\documentclass{pasa}%

\usepackage{graphicx}

\title[]{Coalition Control Model: A Dynamic Resource Distribution Method Based on Model Predicative Control}

\author[]{Weizhi Du$^1$, Harvey Tian$^1$
\affil{$^1$Shanxi Experimental Secondary School}%
}%
\usepackage{float}
\usepackage{aas_macros}
\usepackage{hyperref} 
\usepackage{amsmath}
\usepackage{listings}
\usepackage{mathrsfs}
\usepackage{natbib}
\setcitestyle{square,aysep={},yysep={;},numbers} 

\hypersetup{draft}

\begin{document}
\begin{frontmatter}
\maketitle

\begin{abstract}
Optimization of resource distribution has been a challenging topic in current society. To explore this topic, we develop a Coalition Control Model(CCM) based on the Model Predictive Control(MPC) and test it using a fishing model with linear parameters. The fishing model focuses on the problem of distributing fishing fleets in certain regions to maximize fish caught using either exhaustive or heuristic search. Our method introduces a communication mechanism to allow fishing fleets to merge or split, after which new coalitions can be automatically formed. Having the coalition structure stabilized, the system reaches the equilibrium state through the Nash-Bargaining process. Our experiments on the hypothetical fishing model demonstrated that the CCM can dynamically distribute limited resources in complex scenarios. 
\end{abstract}

\end{frontmatter}

\section{INTRODUCTION }
\label{sec:intro}

How to effectively allocate limited resources while achieving maximized interests has been bewildering people for centuries. Optimization of the resource distribution can be described as finding the optimal solution for a multi-agent control system. In such a system, we first need to evaluate which is more important: maximizing the individual interest or the collective interest, and then to find an appropriate trade-off mechanism when conflicts occur. In some cases that agents can be cooperative, we need to consider the fact that agents can form coalitions to increase their interests, and, possibly, the interest of the entire group. The coalition structure can be used to describe the cooperation among agents. The process of finding the best coalition structure is what we called coalition control\cite{optimal_control}. 


There have been numerous studies on coalition control among different fields, and most of the studies are based on the outcomes from Model Predictive Control(MPC). MPC is a widely used and handy tool when solving optimization problems. As the name suggests, it can produce predictive results of a particular model for a given time horizon\cite{AFRAM2014343}. We can set the boundary conditions so that the function can approach the desired direction. In total, MPC is a solver that can optimize an objective function with constraints\cite{8870090}. Because of its predicting feature, MPC has often been used in the design of coalition control structure. For example, Fele proposed a coalition control system for the optimization of an irrigation canal\cite{FELE2014314}. Masero found a trade-off between energy consumption and quality of service in a large-scale heterogeneous cellular network\cite{9143643}. In terms of traffic control, Asadi optimized the travel time on highway using a coalitional control MPC system\cite{9102355}, and Chanfreut built a model that uses the information of upcoming traffic signals to reduce wait time at stop signs\cite{5454336}. These models have non-linear constraints so the optimization problems are difficult to solve. Therefore, after exploring the fundamental properties of MPC, we have decided to develop a simple mechanism to study coalition control methods. We hope that people who have weak background in optimization and computer algorithm can comprehend our method and apply it to real problems. \\
\indent In this paper, we construct a fishing model to illustrate the idea of coalition control based on MPC. The fishing model includes a certain number of fishing fleets and regions. Each fleet has effort parameters that correspond to how much work it does in each region. Also, we include the fish inflow parameter, which is the amount of fish that come in from other places, and the fish population survival rate, which is the proportion of fish that will still be alive after some time. To make the problem easier to implement, we set linear parameters. In our design, we introduce the ideas of cross-coalition communication and no cross-coalition communication\cite{baldiviesomonasterios2020coalitional}. If the agents are in a no cross-coalition communication system, they have no information regarding how other coalitions behave. We implemented two of such systems, grand coalition and isolated coalition (singleton coalition in some literature). In a cross-coalition system, agents are free to communicate with others and share all the information. We call the coalition structure of such a system controlled coalition. 

Next, our interest is to develop an algorithm that helps the agents make the most appropriate decisions after acquiring others' information. The main decisions to make are whether to merge with other agents or to split from the current coalition\cite{9029893}. As a result, the coalition structure of the agents constantly changes with time. Besides, when designing the algorithm, we need to consider the fact that an individual agent's interest can be harmed after joining a coalition. In this case, we have designed a compensation mechanism for this agent through the redistribution of fish caught. We came up with this specific compensation mechanism after thinking about many real cases in which cooperation is hindered by unbalanced profit distribution. On the other hand, we allow the merge operation to happen unless all the agents are satisfied. In this case, fish redistribution is not necessary. We consider these two scenarios separately, and we call them controlled coalitions with and without fish redistribution.

The proposed method above finds the globally optimal coalition structure (from an individual agent perspective): it calculates the fish caught of every possible coalition structure and finds the maximum value. However, for this method, it can be extremely inefficient in terms of computing time. As a result, we cannot increase the number of fishing fleets and regions or introduce other complicated parameters and mechanisms. Instead, we studied and implemented the method of hierarchical clustering with convex relaxation\cite{10.5555/3104482.3104576}. We applied this method as a heuristic approach to the coalition control problem. This method recursively joins or splits the boats with similar fishing schedules to produce a locally optimal coalition structure. The heuristic approach can reduce computing time significantly, and we can explore problems with large sample sizes and include additional constraints to make the model more realistic.

It is worth to mention that the methods of finding the most appropriate coalition structure include the knowledge of game theory. Some principles can guide us when designing the protocols of a multi-agent system. The first one is social optimum, which describes the circumstance that the collective interest is maximized. For the social optimum solution, we do not allow one agent to know the decisions from others, but we find the optimal solution from a global perspective. The coalition structure from the social optimum solution is a grand coalition\cite{soicalop}. The second one is the Nash-bargaining solution, and in this case, the agents frequently communicate with each other. During the communication, one agent can achieve its best interest after judging the information from the surroundings\cite{8990420}\cite{8865184}. When no one's interest can be further increased, the equilibrium is established. The coalition structure from the Nash-bargaining solution is a controlled coalition\cite{economiclmpc}\cite{coalstruct}. 

In our design, we use MPC to give each boat the capability of predicting the future and finding the best strategy at the current time step. In other words, MPC tells the captain whether to merge with other boats to form a coalition or split from the existing one. The MPC solver is implemented in the MATLAB toolbox (built-in fmincon function), and we take it as part of our design\cite{Mayne2014}.

The unique contributions that distinguish the proposed work from existing researches are threefold: 1) our work develops coalition control methods(globally optimal and heuristic) that can solve resource distribution problems; and 2) we introduce an algorithm that incorporate cross-coalition communication mechanism into the fishing model; and 3) the experimental results of fishing model verified that our proposed methods are viable.

The remainder of the paper is organized as follows. In Section 2, we describe the coalition control model. In Section 3, we present the algorithm of the model. In Section 4, we demonstrate the results from the fishing model to show the validity of our method. This paper is concluded with a summary in Section 5.

\section{Coalition Control Model}
\label{sec:theory}

\subsection{Problem Statement}

We assume that there are $N$ regions where a fishing boat can invest time in fishing. These are identified with the set of numbers $\mathcal{N}:= \left\{ 1,2,...,N \right\}$. We initially assume that populations of fish in the $i$th region, denoted by $x_i(t)$ evolve according to a linear equation of the following kind:
\begin{equation}
x_{i}(t+1)=A_{i}+B_{i} x_{i}(t),
\label{eq:1}
\end{equation}
in the absence of fishing boats. Notice that $A_i$ is a constant inflow of fish, while $B_i$ represents the population's survival rate. This is a simplistic model but still allows us to capture some of the trade-offs involved in coordinating multiple fishing boats. If the geographical distribution of the regions is known and one wants to consider the fish's ability to move from one region to another, the diffusive coupling could be added to the equation.

There are $K$ fishing boats operating in the $N$ regions and each boat $k\in \mathcal{K} := \left\{ 1, 2, ..., K \right\}$ will decide the effort to be devoted at fishing in the region $i$ at time $t$, denoted by $e_{k,i}(t)$. It is assumed that each boat will extract from the corresponding region an amount of fish proportional to the effort $e_{k,i}(t)$ and to the amount of fish $x_i(t)$ in the corresponding region. In particular, then, the evolution of fish can be modeled as:
\begin{equation}
x_{i}(t+1)=A_{i} + B_{i} x_{i}(t) -\sum_{k} \gamma_{k} e_{k, i}(t) x_{i}(t),
\label{eq:2}
\end{equation}
where $\gamma_k$ is the proportionality constant, which may be different for each boat, taking into account the fact that their technologies for fishing might be more or less efficient. Each boat $k$ is interested in maximizing its caught:
\begin{equation}
O_{k}=\sum_{t} \sum_{i} e_{k, i}(t) x_{i}(t),
\label{eq:3}
\end{equation}
over a certain time interval $\left\{0, 1, 2, ..., T \right\}$ that will be used as a prediction horizon of our coalitional MPC model. Each boat's effort is supposed to be non-negative, and the total effort in the various areas needs to add up to 1,
\begin{equation}
e_{k, i}(t) \geq 0, \quad \sum_{i} e_{k, i}(t)=1.
\label{eq:5}
\end{equation}

Therefore, each boat can choose where it wants to focus its fishing efforts and determine when and where to fish. Notice that congestion in a particular area rich in fish will deplete the area and lead to a loss of fishing profitability.

Fishing boats are allowed to form coalitions. A coalition is a subset $ \mathcal{C} \in \mathcal{K}$ of boats, which jointly optimize their efforts to maximize the total amount of fish they can extract at sea. In particular, the objective function is:
\begin{equation}
O_{\mathcal{C}_l}=\sum_{t} \sum_{k\in \mathcal{C}_l} \sum_{i} \gamma_{k} e_{k, i}(t) x_{i}(t).
\label{eq:4}
\end{equation}

Notice that a coalition could be a single boat; in which case $\mathcal{C} = \left\{ k \right\}$ and consistently we see that $O_k = O_{\left\{ k \right\} }$. At each time $t$ the fleet of boats is partitioned in a number of coalitions:
\begin{equation}
\mathcal{C}_{\mathcal{K}}(t)=\mathcal{C}_{1}(t) \cup \mathcal{C}_{2}(t) \cup \ldots \cup \mathcal{C}_c(t),
\end{equation}
where $\mathcal{C}_{\mathcal{K}}(t)$ is the coalition structure at time $t$ (time could change), $c$ is the total number of coalitions in $\mathcal{C}_{\mathcal{K}}$ and $\mathcal{C}_l (t)$ denotes the $l$th coalition. Each boat only belongs to one given coalition at any time $t$. At each time $t$ each coalition solves the following optimization problem of maximizing objective function $O_{\mathcal{C}_{\mathcal{K}}}$:
\begin{equation}
\begin{split}
\max_{\mathcal{C}_{\mathcal{K}} \in \mathcal{P}} &\ O_{\mathcal{C}_{\mathcal{K}}}\\
s.t. &\ \sum_{i\in \mathcal{N}} e_{k, i}(t)=1,e_{k, i}(t) \geq 0,  k \in \mathcal{K}\\      
     &\ e_{k, i}= e_{j, i}, k,j \in \mathcal{C}_l,\mathcal{C}_l\in \mathcal{C_\mathcal{K}},i\in\mathcal{N} \\
     &\ x_{\mathcal{C}}^{eq}-\mathcal{R}< x_{\mathcal{C}}< x_{\mathcal{C}}^{eq}+\mathcal{R}
\end{split}
\label{eqob1}
\end{equation}
where $\mathcal{P}$ is the set for entire coalition structures, $x_{\mathcal{C}}^{eq}$ is using the equilibrium result from MPC to design a sustainable constraint so that the total amount of fish can be maintained at a  certain level, $\mathcal{R}$ is the relaxation for tolerance. $e_{k, i}= e_{j, i}$ means that when two boats form a coalition, they invest same amount of effort at region $i$.

\subsection{Evolution of Coalition Structure}

As mentioned earlier, we assume that the evolution of the coalition structure has the following different designs:
\begin{enumerate}%
\item No cross-coalition communication: this is in the absence of information on how other coalitions will behave. It is an optimistic view that other agents' efforts are 0. In terms of coalition structure, we have a grand coalition in which the entire group's interest matters the most. In our model, all fishing fleets form one coalition, and the total fish caught is maximized. Also, we have an isolated coalition, in which each agent ignores the intentions and decisions of other agents. 
\item Cross-coalition communication: in this case the efforts $e_{k,i}(t)$ are fixed and a-priori communicated. This communication is done only once every time-step by broadcasting their fishing schedule to all other coalitions. The first coalition to do so will typically use the predicted schedules to build its fishing schedule from the previous time. After the communication is done, the agent chooses to merge with or split from other agents to harvest more.
\end{enumerate}

When designing the mechanisms of merging and splitting, we need to consider the possibility that one agent/coalition's interest may be jeopardized after joining a new coalition, even though the new coalition's interest is increased. At this point, if we decide the merge operation continues to happen, we need to redistribute the fish caught to compensate for the unhappy agents. We will allow the merge operation to happen only if both agents(or coalitions) are benefited at the same time. In this case, we do not need to redistribute the fish caught. To formulate, we now carry out an optimization of the objective function $O_{\mathcal{C}_{m} \cup \mathcal{C}_{n}}$, where $\mathcal{C}_{m}$ and $\mathcal{C}_{n}$ are the two coalitions considering a merge operation. The two criteria of a merging operation can be shown as:
\begin{itemize}
    \item With fish redistribution
    \begin{align}
        & \text{Merging:} \quad O_{\mathcal{C}_{m} \cup \mathcal{C}_{n}}^{*} \geq O_{\mathcal{C}_{m}}^{*}+O_{\mathcal{C}_{n}}^{*}  \label{eq:m1} \\
        & \text{Splitting:} \quad O_{\mathcal{C}_{m} \cup \mathcal{C}_{n}}^{*} < O_{\mathcal{C}_{m}}^{*}+O_{\mathcal{C}_{n}}^{*}. \label{eq:s1}
    \end{align}
    We denote $O^{*}$ the optimal fish caught, and it comes from the predicted results of MPC. The condition implies that the predicted optimal total amount of fish of the merged coalition is higher than the sum of the fish caught by individual coalitions acting alone. It is agreed that coalition $\mathcal{C}_{m}$, denoted by $O^{*}_{\mathcal{C}_{m} \cup \mathcal{C}_{n} | \mathcal{C}_{m}}$, will get an amount of fish equals to:
    \begin{equation}
        O^{*}_{\mathcal{C}_{m} \cup \mathcal{C}_{n} | \mathcal{C}_{m}} =  O_{\mathcal{C}_{m} \cup \mathcal{C}_{n}}^{*} \frac{O_{\mathcal{C}_{m}}^{*}}{O_{\mathcal{C}_{m}}^{*}+O_{\mathcal{C}_{n}}^{*}},
    \end{equation}
    and coalition $\mathcal{C}_{n}$, denoted by $O^{*}_{\mathcal{C}_{m} \cup \mathcal{C}_{n} | \mathcal{C}_{n}}$, will get an amount of fish equals to:
        \begin{equation}
        O^{*}_{\mathcal{C}_{m} \cup \mathcal{C}_{n} | \mathcal{C}_{n}}  =  O_{\mathcal{C}_{m} \cup \mathcal{C}_{n}}^{*} \frac{O_{\mathcal{C}_{n}}^{*}}{O_{\mathcal{C}_{m}}^{*}+O_{\mathcal{C}_{n}}^{*}}.
    \end{equation}
    Once a coalition is formed which entails fish redistribution, the ratios $r_m :=\frac{O_{\mathcal{C}_{m}}^{*}}{O_{\mathcal{C}_{m}}^{*}+O_{\mathcal{C}_{n}}^{*}}$ and $r_n := \frac{O_{\mathcal{C}_{n}}^{*}}{O_{\mathcal{C}_{m}}^{*}+O_{\mathcal{C}_{n}}^{*}}$ are recorded and used for subsequent attribution of the amount of fish for each coalition. 
    
    The splitting condition suggests that if the predicted sum of the fish caught by individual coalitions acting alone is greater than the total amount of fish of the merged coalition, coalition $\mathcal{C}_{m} \cup \mathcal{C}_{n}$ breaks up into $\mathcal{C}_{m}$ and $\mathcal{C}_{n}$. 
    \item Without fish redistribution
    \begin{equation}
        \begin{aligned}
             \text{Merging:} \quad & O^{*}_{\mathcal{C}_{m} \cup \mathcal{C}_{n} | \mathcal{C}_{m}}  \geq O_{\mathcal{C}_{m}}^{*} \quad \text{and} \\
             & O^{*}_{\mathcal{C}_{m} \cup \mathcal{C}_{n} | \mathcal{C}_{n}}  \geq O_{\mathcal{C}_{n}}^{*}
        \end{aligned}
        \label{eq:m2}
    \end{equation}
    \begin{equation}
        \begin{aligned}
             \text{Splitting:} \quad & O^{*}_{\mathcal{C}_{m} \cup \mathcal{C}_{n} | \mathcal{C}_{m}}  < O_{\mathcal{C}_{m}}^{*} \quad \text{or} \\
             & O^{*}_{\mathcal{C}_{m} \cup \mathcal{C}_{n} | \mathcal{C}_{n}}  < O_{\mathcal{C}_{n}}^{*}.
        \end{aligned}
        \label{eq:s2}
    \end{equation}    
    Notice that the conditions above have to be satisfied at the same time to complete a merge operation. However, if any of the coalition is not satisfied with the current fish caught, the coalition can choose to leave the currently merged coalition. In this case, it is not necessary to redistribute the fish, and coalition $\mathcal{C}_{m}$ has the amount of fish,
    \begin{equation}
        O^{*}_{\mathcal{C}_{m} \cup \mathcal{C}_{n} | \mathcal{C}_{m}} = \sum_{k \in \mathcal{C}_{m}} \sum_{i} e_{k, i}^{*}(t) x_{i}^{*}(t).
    \end{equation}
    We denote by $u^{*}$ and $x^{*}$ the optimal schedule and state-solution corresponding to the merged coalition. A similar expression holds for $\mathcal{C}_{n}$.
\end{itemize}


At each new time instant $t$, the coalitions coming from the previous time $t-1$, will check if merge or split operations will occur and decide the new coalition structure at time $t$. Sometimes one could argue that this operation is only carried out every $T$ time steps or longer to avoid coalition forming and splitting, which is too fast. Once the coalition structure at the current time has been agreed upon, then the optimization is carried out sequentially, and optimal fishing efforts are computed. In total, we let the fishing fleets choose to merge or split every time step $T$. Then, the fish stock evolves in time, and the problem is reformulated at the following time instant, giving rise to a coalitional receding horizon control protocol.

Here, we provide a brief time complexity analysis for our method. Given $N$ regions and $K$ boats, we need to consider the coalition structure for all possible combinations, including a coalition consisting two boats, three boats, etc.. It means that we have to conduct an exhaustive search for all the coalitions. For each coalition structure, we call MPC to make decisions about whether to merge or not. For a coalition with two boats, the number of coalitions is $C_K^2$, and the total number of coalition we need to consider for time $t=1$ becomes, $C_K^2+C_K^3+C_K^4+\cdots+C^K_K = 2^K-K-1$. Notice that this is the maximum amount of coalition possibilities at one time step. Then for the entire time, we have time complexity in the form of $\mathcal{O}(2^K)$, which is exponential with increasing number of boat. Also, the MPC run time increases dramatically with a large number of $N$ and $K$, so it takes extremely long time if we want to include more boats and regions.

\subsection{Accelerated Coalition Control Model}
To improve the computing efficiency, we studied the hierarchical clustering algorithm and applied it to our fishing problem. The major difference is that our algorithm finds the global optimal coalition structure, whereas the hierarchical clustering algorithm is heuristic and finds the local optimum. Here, the terms cluster and coalition are equivalent. In terms of the merging process, we have each agent to start in its own cluster, and the cluster chooses to merge based on the similarity between different clusters. We measure the similarity by constructing virtual effort vectors $\mathbf{v}_k$ for boat $k$ and then calculating the Euclidean distance between virtual effort vectors. We initialize the virtual effort vectors the same as the real effort vectors. If the virtual effort vectors are close enough, pairs of clusters are merged as one moves up the hierarchy. Therefore, we adjust the objective function as following:
\begin{equation}
\begin{split}
\min_{\mathbf{E},\mathbf{V}} &-\ O_{\mathbf{E}}+\mu \sum_{k}\|\textbf{e}_k-\textbf{v}_k\|_2^2+ \gamma \sum_{1\leq k < l\leq K}\|\textbf{v}_k-\textbf{v}_l\|_2^2 \\
s.t. &\ \sum_{i} e_{k, i}(t)=1,e_{k, i}(t) \geq 0,  k \in \mathcal{K}, i \in \mathcal{N}\\
     &\ x_{\mathcal{C}}^{eq}-\mathcal{R}< x_{\mathcal{C}}< x_{\mathcal{C}}^{eq}+\mathcal{R},
     \end{split}
\end{equation}
where $\mathbf{e}_k=(e_{k,1}, e_{k,2},\dots,e_{k,N})$ is the real effort vector of the $kth$ ship in different regions, $\mathbf{E}=(\textbf{e}_1,\textbf{e}_2,\dots,\textbf{e}_K)$ is the matrix saving effort vectors for all boats, , $\textbf{v}_k=(v_{k,1},v_{k,2},\dots,v_{k,N})$ is the virtual effort vector and it reflects the coalition structure, i.e. when $\textbf{v}_k=\textbf{v}_l$, it means that $k,l$ belongs to the same coalition, and $\mu, \gamma$ are the trade-off parameters. $\mu$ determines how much difference is allowed between virtual and real effort vectors. $\gamma$ determines how fast the coalition structure can be changed. For instance, when $\gamma=1$, we introduce another threshold parameter $\delta$ so that when $\|\textbf{v}_i-\textbf{v}_j\|\leq \delta$, $\textbf{v}_i$ and $\textbf{v}_j$ will be merged into a point, and the corresponding boat will form a coalition. Then if $\gamma$ is significantly greater than 1, the merging process completes quickly, i.e. multiple coalitions form in a short time. If $\gamma$ is less than one, the coalition will consider splitting. As a result, we adjust the $\gamma$ factor for merging and splitting, and we will focus on merging below. After establishing the objective function, we use the fixed-variable method to solve real effort matrix $\mathbf{E}$ and virtual coalition structure $\mathbf{V}$ alternately. The time complexity for the heuristic approach is $\mathcal{O}(K^2)$.

\begin{enumerate}
\item Solving $\mathbf{V}$ with fixed $\mathbf{E}$ \\
First, we have the information of coalition structure or cluster $\mathbf{V}$ and the real effort vectors $\mathbf{E}$. Then we need to solve for the virtual effort vector $\mathbf{v}$, and the coalition structure will be updated as $\mathbf{V}$. The objective can be adjusted as:
\begin{equation}
\begin{split}
\min_{\mathbf{V}} & L(\mathbf{V}) = \mu \sum_{k}\|\textbf{e}_k-\textbf{v}_k\|_2^2+ \gamma \sum_{1\leq k < l \leq K}\|\textbf{v}_k-\textbf{v}_l\|_2^2 \\
s.t. &\ \sum_{i} e_{k, i}(t)=1,e_{k, i}(t) \geq 0,  k \in \mathcal{K}, i \in \mathcal{N}\\
     &\ x_{\mathcal{C}}^{eq}-\mathcal{R}< x_{\mathcal{C}}< x_{\mathcal{C}}^{eq}+\mathcal{R}
     \label{eqob2}
     \end{split}
\end{equation}
Let $\mathcal{V}_k=\{\mathbf{v}_l|\mathbf{v}_k\neq \mathbf{v}_l,(\mathbf{v}_k,\mathbf{v}_l)\in \mathbf{V}\}$, we solve for optimal $\mathbf{V}$ for all $k\in \mathcal{N}$ by setting the gradient of Eq.\ref{eqob2} to be 0:
\begin{equation}
\frac{\partial L(v)}{\partial v_k} = 2\mu(\textbf{e}_k-\textbf{v}_k)+2\gamma \sum_{l\in \mathcal{V}_k}(\textbf{v}_k-\textbf{v}_l) =0.
\end{equation}

Then, we can get the virtual effort vector as:
\begin{equation}
    \mathbf{v}_k=\frac{1}{\gamma|\mathcal{V}_k|-\mu}(\gamma \sum_{l\in \mathcal{V}_k}\mathbf{v}_l-\mu \mathbf{e}_k).
\end{equation}

\item Solving $\mathbf{E}$ with fixed $\mathbf{V}$ \\
When we have fixed $\mathbf{V}$, the objective function becomes:
\begin{equation}
\begin{split}
\min_{\mathbf{E}} &-\ O_{\mathbf{E}}+\mu \sum_{k}\|\textbf{e}_k-\textbf{v}_k\|_2^2\\
s.t. &\ \sum_{i} e_{k, i}(t)=1,e_{k, i}(t) \geq 0,  k \in \mathcal{K}, i \in \mathcal{N}\\
     &\ x_{\mathcal{C}}^{eq}-\mathcal{R}< x_{\mathcal{C}}< x_{\mathcal{C}}^{eq}+\mathcal{R}.
\end{split}
\end{equation}
At this point, we solve the objective function using MPC similarly as in our previous method (see Eq.\ref{eqob1}), only with the difference that the objective function becomes $-O_{\mathbf{E}}+\mu \sum_{k}\|\textbf{e}_k-\textbf{v}_k\|_2^2$.
\end{enumerate}

\section{ALGORITHM}

\begin{figure}[ht]
\centering
\includegraphics[width=\columnwidth]{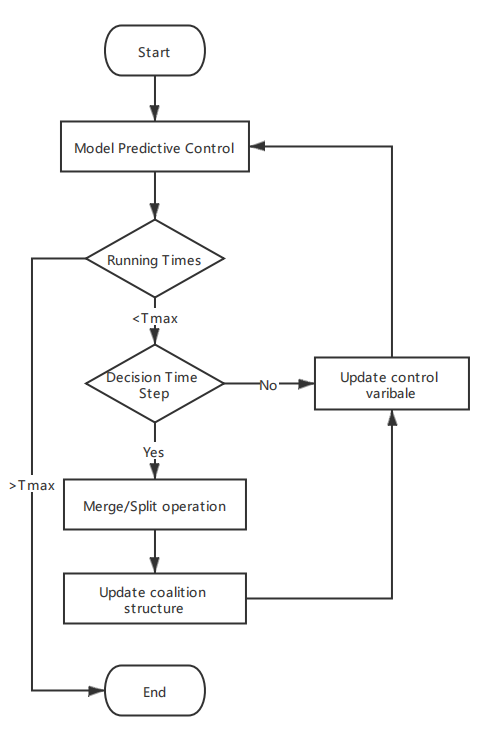}
\caption{Flowchart of cross-coalition communication algorithm for a multi-agent system.}
\label{fig:closedcross}
\end{figure}

We provide a flow chart for cross-coalition communication systems to demonstrate the proposed algorithm in Fig. \ref{fig:closedcross}. We start from the multi-agent system node, which includes the information of coalition structure, state variables, and restraints. Until the iteration is terminated, the program will call the MPC solver to find each coalition's optimal schedule. We set the prediction horizon to be long enough so the system can reach equilibrium in MPC. If it is the time step we need to decide about adjusting the coalition structure, we use the previous MPC results to merge or split coalitions. If not, we employ the MPC solver results to update variables without altering the coalition structure. The system's outputs are the values of control variables, objective function, individual interest, and collective interest.\\ 
\indent Next, we illustrate the pseudo-code from Listing \ref{lst:MPC} for our algorithm of finding optimal solutions from MPC. When implementing, we choose the input to be the known conditions about the entire system, which include the fish population $\mathbf{x}$, initial guess of efforts $\hat{\mathbf{E}}$, temporary efforts $\Tilde{\mathbf{E}}$, number of coalitions $n$, maximum iteration $max$ and optimal problem 
$\mathcal{O}_{\mathcal{K}}^{*}$. So the optimization problem can be formulated and solved using

\begin{lstlisting}[caption=Pseudo-code illustration of finding optimal solutions using MPC,mathescape=true,label=lst:besteq,breaklines=true,basewidth=4pt,prebreak=,postbreak=,frame=single,label=lst:MPC]
Inputs: fish population $\mathbf{x}$, initial guess of efforts $\hat{\mathbf{E}}$, temporary efforts $\Tilde{\mathbf{E}}$, number of coalitions $c$, maximum iteration $max$ and optimal problem $\mathcal{O}_{C_\mathcal{K}}^{*}$

if $c$ = 1   % grand coalition
    max = 1
end if
    
$\mathbf{x}^{eq} = \mathbf{x}$

for it = 1 to max do
    for $i$ = 1 to $c$ do
    
        $\mathcal{O}_{\mathcal{K}}^{*}\gets$ solve optimal problem w.r.t $\Tilde{\mathbf{E}}_{ \mathcal{C}_{i}}$ and $\mathbf{x}^{eq}$ by using $\hat{\mathbf{E}}_{ \mathcal{C}_{i}}$ $\mathbf{x}^{eq}$ as the initial guess
        
        $\Tilde{\mathbf{E}}_{ \mathcal{C}_{i}}$ and $\hat{\mathbf{E}}_{ \mathcal{C}_{i}} \gets \mathbf{E}^{*}_{ \mathcal{C}_{i}}$; $\mathbf{x}^{eq}\gets \mathbf{x}^{eq*}$
    end for
end for
$\mathbf{E}^{eq}$ = $\left[\begin{array}{c}\Tilde{\mathbf{e}}_{ \mathcal{C}_1}\\ \Tilde{\mathbf{e}}_{ \mathcal{C}_2}\\ \vdots\\ \Tilde{\mathbf{e}}_{ \mathcal{C}_c}\end{array}\right] \triangleq \left[\begin{array}{c}\mathbf{e}^{eq}_{ \mathcal{C}_1}\\ \mathbf{e}^{eq}_{ \mathcal{C}_2}\\ \vdots\\ \mathbf{e}^{eq}_{ \mathcal{C}_c}\end{array}\right]$ 

Output:vector $\mathbf{E}^{eq}$, $\mathbf{x}^{eq}$

\end{lstlisting}

\noindent MPC. The control variables of each agent can be updated and reused as the inputs in the next run. After running the MPC for the maximum number of iterations, we can find optimal solutions for the given coalition. Generally, the optimal solutions are in equilibrium states. The outputs are the solutions, which include the effort parameters of all the sub-coalitions $\mathbf{E}^{eq}$, and equilibrium fish population in each region $\mathbf{x}^{eq}$. Notice that in the algorithm, the MPC solver is in the center of the nested loops, and it takes great computational effort for this algorithm.

Based on the information from MPC, we have designed a method to alter the coalition structure. First, we consider when two coalitions intend to merge. The merging conditions are mentioned in Eq.(\ref{eq:m1}) and Eq.(\ref{eq:m2}). In Listing \ref{lst:merge}, we present the pseudo-code for merging operation, which takes inputs of coalition structure $\mathcal{C}_{\mathcal{K}}$, number of coalitions $n$, condition of merging $\mathscr{M}$. We set a temporary coalition structure $\mathcal{C}_{\mathcal{K}}^{temp}$, a new coalition after merging $\mathcal{C}_{\mathcal{K}}^{new}$ inside the loop to exchange coalition structure information. In the end, the algorithm produces an updated coalition structure $\mathcal{C}_{\mathcal{K}}^{*}$.

Similarly, we present the pseudo-code for splitting operation in Listing \ref{lst:split}. The conditions for splitting are mentioned in Eq.(\ref{eq:s1}) and Eq.(\ref{eq:s2}).
\newline 

\begin{lstlisting}[caption=Pseudo-code illustration of merge operation,mathescape=true,label=lst:Merge,breaklines=true,basewidth=4pt,prebreak=,postbreak=,frame=single,label=lst:merge]
Inputs: coalition structure $\mathcal{C}_{\mathcal{K}}$, number of coalitions $c$, condition of merging $\mathscr{M}$

if $c$ >1
    $\mathcal{C}_{\mathcal{K}}^{temp} = \mathcal{C}_{\mathcal{K}}$
    for $i$ = 1 to $c-1$ do
        for $j$ = $i + 1$ to $c$ do
            $\mathcal{C}_{\mathcal{K}}^{new} \gets \text{ merging } \mathcal{C}_{i}\text{ and } \mathcal{C}_{j}\text{ in }\mathcal{C}_{\mathcal{K}}^{temp}$
        
            Call algorithm from Listing $\ref{lst:MPC}$ for $\mathcal{C}_{\mathcal{K}}^{new}$ and $\mathcal{C}_{\mathcal{K}}^{temp}$
        
            if $\mathscr{M}$ meets  
                $\mathcal{C}_{\mathcal{K}}^{temp} = \mathcal{C}_{\mathcal{K}}^{new}$
            end if
        end for
    end for
    $\mathcal{C}_{\mathcal{K}}^{*} = \mathcal{C}_{\mathcal{K}}^{temp}$
else
    $\mathcal{C}_{\mathcal{K}}^{*} = \mathcal{C}_{\mathcal{K}}$
end if

Output:matrix $\mathcal{C}_{\mathcal{K}}^{*}$
\end{lstlisting}

\begin{lstlisting}[caption=Pseudo-code illustration of split operation,mathescape=true,label=lst:Split,breaklines=true,basewidth=4pt,prebreak=,postbreak=,frame=single,label=lst:split]
Input: coalition structure $\mathcal{C}_{\mathcal{K}}$, number of coalitions $c$, condition of splitting $\mathscr{S}$

$\mathcal{C}_{\mathcal{K}}^{temp} = \mathcal{C}_{\mathcal{K}}$
for ${i}$ = 1 to $c-1$ do
    if $c(\mathcal{C}_{i}) > 2$
        for $j = 1$ to $c(\mathcal{C}_{i})$
            $\mathcal{C}_{\mathcal{K}}^{new} \gets \text{ split } 
            \mathcal{C}_{j}\text{ in }\mathcal{C}_{\mathcal{K}}^{temp} $

            Call algorithm from Listing $\ref{lst:MPC}$ for $\mathcal{C}_{\mathcal{K}}^{new}$ and $\mathcal{C}_{\mathcal{K}}^{temp}$
        
            if $\mathscr{S}$ meets  
                $\mathcal{C}_{\mathcal{K}}^{temp} = \mathcal{C}_{\mathcal{K}}^{new}$
            end if
        end for
    end if
end for
$\mathcal{C}_{\mathcal{K}}^{*} = \mathcal{C}_{\mathcal{K}}^{temp}$
    
Output:matrix $\mathcal{C}_{\mathcal{K}}^{*}$
\end{lstlisting}

At last, we illustrate the pseudo-code for hierarchical clustering method. We set coalition threshold $\epsilon$ to determine whether the boat should merge, This is a bottom-up approach so we need additional trade-off parameters $\mu$ and $\gamma$ to stablize the coalition structure. 
\begin{lstlisting}[caption=Pseudo-code illustration of hierarchical clustering method, mathescape=true,label=lst:Split,breaklines=true,basewidth=4pt,prebreak=,postbreak=,frame=single,label=lst:split]
Input: individual boat efforts $\mathbf{E}$, coalition structure $\mathbf{V}$, number of coalitions $c$, maximum iteration $max$, coalitions thresholds $\epsilon$, trade-off parameters $\mu,\gamma$

for it = 1 to max do
    $\mathbf{V}=\mathbf{E}$
    for ${i}$ = 1 to $c-1$ do
        solve $\mathbf{v}_i=\frac{1}{\gamma|V_k|-\mu}(\gamma \sum_{l\in V_k}\mathbf{v}_l-\mu \mathbf{e}_k$)
    end for
    $\mathbf{V}^{temp}=\mathbf{V}$
    for ${i}$ = 1 to $c-1$ do
        for ${j}$ = ${i} + 1$ to $c$ do
            dist = $\|\mathbf{v}_i-\mathbf{v}_j\|_2^2$
            if dist < $\epsilon$
                $\mathbf{V}^{new} \gets \text{ merging } \mathcal{C}_{i}\text{ and } \mathcal{C}_{j}\text{ in }\mathbf{V}^{temp}$
            end if 
        end for
    end for
    
    Call algorithm from Listing $\ref{lst:MPC}$ for $\mathbf{E}^{new}$ 
    (modify the objective function to $-O_{\mathbf{E}}+\mu \sum_{i}\|\mathbf{e}_i-\mathbf{v}_i\|_2^2$)
        
    if $O_\mathbf{E}$ is not increasing or satisfying specified conditions
        $\mathbf{V}=\mathbf{V}^{new}$
        $\mathbf{E}=\mathbf{E}^{new}$
        break
    end if
end for

Output:matrix $\mathbf{E}$, matrix $\mathbf{V}$
\end{lstlisting}

\section{EXPERIMENTS}

\subsection{Parameters and Setups}
First, we show the initial conditions used in the paper. We construct the problem to have four fishing regions and six fishing fleets. Each region has its fish inflow rate and survival rate. Also, each boat has the fishing capability factor $\gamma$, so we can differentiate. The numeric values of the parameter we used are shown in Table \ref{tab:Par}.
\begin{table}[htbp]
	\centering
	\caption{Parameters we chose to use in order to make the fishing model in a way that the grand coalition is the efficient structure.}
	\begin{tabular}{lll}
		Symbol & Meaning & Value\\
		\hline
		$N$ & number of regions & 4  \\
		$K$ & fishing fleets number  & 6 \\
    	$A_1$ & 1st region fish inflow & 300   \\
		$A_2$ &  2nd region fish inflow  & 450\\
		$A_3$ &  3rd region fish inflow & 350 \\
		$A_4$ &  4th region fish inflow  & 200\\
    	$B_1$ & 1st region survival rate & 0.2  \\
		$B_2$ & 2nd region survival rate & 0.3\\	
		$B_3$ & 3rd region survival rate & 0.45 \\
    	$B_4$ & 4th region survival rate & 0.6  \\
		$\gamma_1$ & 1st boat capability & 0.08 \\	
		$\gamma_2$ & 2nd boat capability  & 0.1\\
    	$\gamma_3$ & 3rd boat capability  & 0.12  \\
		$\gamma_4$ & 4th boat capability  & 0.15 \\
    	$\gamma_5$ & 5th boat capability  & 0.20 \\
    	$\gamma_6$ & 6th boat capability  & 0.28 \\
       	$x_1^0$ & 1st region initial fish & 200 \\
       	$x_2^0$ & 2nd region initial fish & 300 \\
       	$x_3^0$ & 3rd region initial fish & 150 \\
       	$x_4^0$ & 4th region initial fish & 250 \\
	    \hline
	\end{tabular}
	\label{tab:Par}
\end{table}

Under the parameter settings, we conclude that the grand coalition, in which all fishing fleets are distributed to one coalition, can attain the maximum fish caught. The apparent reason is that our model is linear, and the fish inflow is vast. In other words, the fish is abundant, and all fishing fleets do not need to worry too much about not catching anything. As a result, we have a reference point to compare if our result can be close to the theoretically best result. On the other hand, we can also set each boat in its own coalition to eliminate the effect of cooperation(isolated coalition). Also, when running our algorithm of controlled coalition, we find the coalition structure naturally becomes the structure of grand coalition after a certain time period. It is not surprising to find this phenomenon since our model encourages cooperation, but we want to see the coalition structure ends differently. Therefore, we decide that our coalition control method does not allow to converge to a grand coalition, and the maximum number of agents in one coalition is 3. We expect the fish caught less than the result from grand coalition. \\
\indent Besides, we set the total time step to be 720. If each step represents one day, the total fishing period is two years. Also, we decide that the fleet will decide to merge or split each month so that the computation can be efficient. In other words, the coalition structure will be adjusted in 30-time steps.   

\subsection{Results}
\subsubsection{Coalition Control Model}
After running the implemented algorithm, we have the following results.
\begin{figure}[H]
\begin{center}
\includegraphics[width=\columnwidth]{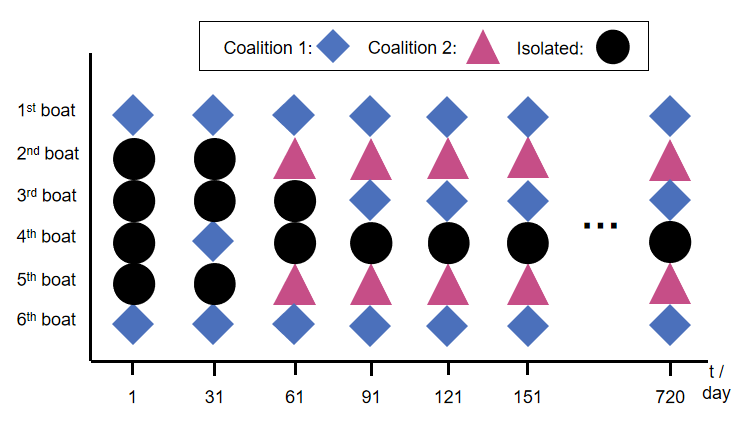}
\caption{How coalition structure changes after each month with fish redistribution. After 90 days, we find that the coalition structure remains the same, which suggests that the dynamic problem reaches a steady state.}
\label{fig:dis}
\end{center}
\end{figure}

\begin{figure}[H]
\begin{center}
\includegraphics[width=\columnwidth]{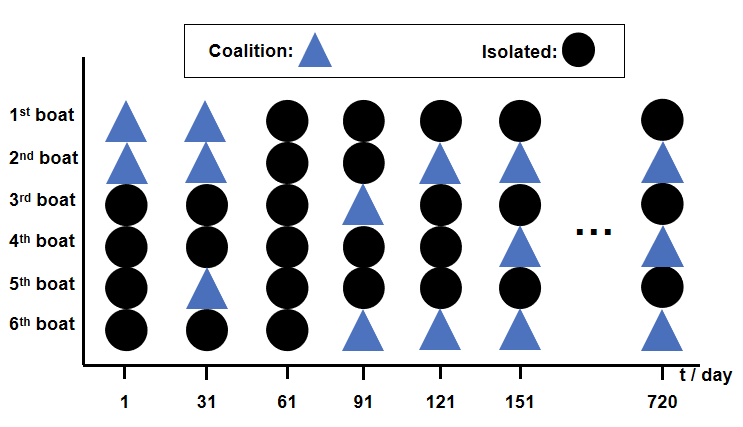}
\caption{How coalition structure changes after each month without fish redistribution. If without fish redistribution, the coalition structure reaches steady-state at 150 day. It takes longer time comparing to the case with fish redistribution.}
\label{fig:nodis}
\end{center}
\end{figure}

First, we show how the coalition structure changes in our controlled coalition method with and without fish redistribution as shown Fig. \ref{fig:dis} and Fig. \ref{fig:nodis}. We observe that the coalition structures both reach a steady-state, and it means that there has to be a dynamic balance between total fish caught and total fish increased. So the sustainability of the fish population is preserved, yet we can maximize the profit from fishing. For the two merging and splitting conditions, we see the finalized coalition structures to be different. For the controlled coalition with the fish redistribution, we see that it takes 90 days to reach a steady structure, and 5 boats merge into 2 coalitions. For the controlled coalition without fish redistribution, the coalition structure changes more vibrantly in the first 150 days until it reaches an equilibrium. Moreover, in the end, only 3 boats decide to cooperate, and they merge into one single coalition. This is reasonable because of the condition without fish redistribution is more restricted than the other condition. 
\begin{figure}[H]
\begin{center}
\includegraphics[width=\columnwidth]{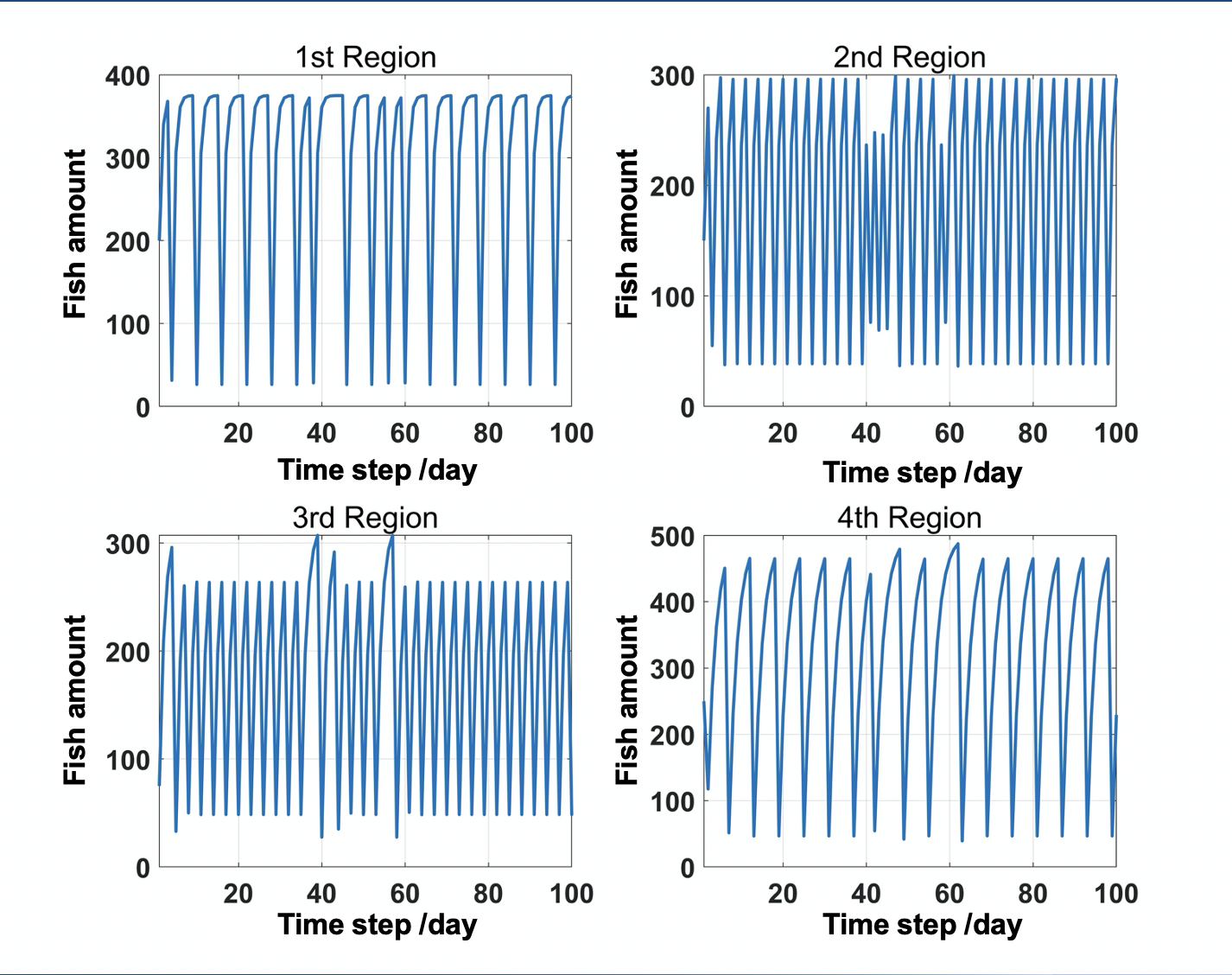}
\caption{The amount of fish changes in each time step from start to day 100. We see the periodic behavior changes with the change of coalition structure.}
\label{fig:condition}
\end{center}
\end{figure}
Next, we show how the fish amount varies in each region in Fig. \ref{fig:condition}. We observe periodic behaviors of the fish amount in all regions. It suggests that fish from other places fill in quickly once the fish in a region is depleted because of the massive inflow. Another conclusion we can draw from the figure is that the periodic behavior changes along with the change of coalition structure, even though the time step may not perfectly match. From Fig. \ref{fig:condition}, we can also conclude that once a region has a large fish amount, the fishing fleets come and harvest quickly. Then the fleets pay less attention to this region for 2-4 days until the fish become crowded again. In the end, the fish removed (by nature and fishing fleets) and fish gained (from other places) are in a dynamic equilibrium, which also suggests that the coalition structure is stabilized when the equilibrium is reached.  

\begin{table}[ht]
	\centering
	\caption{Grand Coalition: The total fish caught of entire fleet and each boat in 2 years/720 time steps.}
	\begin{tabular}{lll}
		Symbol & Description & Fish Value\\
		\hline
		$\mathcal{F}$ & total fish caught & $343.25 \times 10^{3}$  \\
		$\mathcal{F}_1$ & 1st boat & $29.53 \times 10^{3}$  \\	
		$\mathcal{F}_2$ & 2nd boat & $36.91 \times 10^{3}$ \\
    	$\mathcal{F}_3$ & 3rd boat & $44.29 \times 10^{3}$   \\
		$\mathcal{F}_4$ & 4th boat & $55.36 \times 10^{3}$  \\
    	$\mathcal{F}_5$ & 5th boat & $72.82 \times 10^{3}$  \\
    	$\mathcal{F}_6$ & 6th boat & $103.34 \times 10^{3}$  \\
	    \hline
	\end{tabular}
	
	\label{tab:fishtoatl}
\end{table}

 \begin{table}[ht]
	\centering
	\caption{Isolated Coalition: The total fish caught of entire fleet and each boat in 2 years/720 time steps.}
	\begin{tabular}{lll}
		Symbol & Description & Fish Value\\
		\hline
		$\mathcal{F}$ & total fish caught & $337.22 \times 10^{3}$  \\
		$\mathcal{F}_1$ & 1st boat & $29.01 \times 10^{3}$  \\	
		$\mathcal{F}_2$ & 2nd boat & $36.26 \times 10^{3}$ \\
    	$\mathcal{F}_3$ & 3rd boat & $43.51 \times 10^{3}$   \\
		$\mathcal{F}_4$ & 4th boat & $54.39 \times 10^{3}$  \\
    	$\mathcal{F}_5$ & 5th boat & $72.52 \times 10^{3}$  \\
    	$\mathcal{F}_6$ & 6th boat & $101.53 \times 10^{3}$  \\
	    \hline
	\end{tabular}
	\label{tab:fishtoat2}
\end{table}
 \begin{table}[ht]
	\centering
	\caption{Controlled Coalition with fish redistribution: The total fish caught of entire fleet and each boat in 2 years/720 time steps.}
	\begin{tabular}{lll}
		Symbol & Description & Fish Value\\
		\hline
		$\mathcal{F}$ & total fish caught & $343.00 \times 10^{3}$  \\
		$\mathcal{F}_1$ & 1st boat & $29.51 \times 10^{3}$  \\	
		$\mathcal{F}_2$ & 2nd boat & $36.88 \times 10^{3}$ \\
    	$\mathcal{F}_3$ & 3rd boat & $44.26 \times 10^{3}$   \\
		$\mathcal{F}_4$ & 4th boat & $55.32 \times 10^{3}$  \\
    	$\mathcal{F}_5$ & 5th boat & $73.76 \times 10^{3}$  \\
    	$\mathcal{F}_6$ & 6th boat & $103.27 \times 10^{3}$  \\
	    \hline
	\end{tabular}
	\label{tab:fishtoat3}
\end{table}
 \begin{table}[ht]
	\centering
	\caption{Controlled Coalition without fish redistribution: The total fish caught of entire fleet and each boat in 2 years/720 time steps.}
	\begin{tabular}{lll}
		Symbol & Description & Fish Value\\
		\hline
		$\mathcal{F}$ & total fish caught & $342.84 \times 10^{3}$  \\
		$\mathcal{F}_1$ & 1st boat & $29.49 \times 10^{3}$  \\	
		$\mathcal{F}_2$ & 2nd boat & $36.86 \times 10^{3}$ \\
    	$\mathcal{F}_3$ & 3rd boat & $44.24 \times 10^{3}$   \\
		$\mathcal{F}_4$ & 4th boat & $55.30 \times 10^{3}$  \\
    	$\mathcal{F}_5$ & 5th boat & $73.73 \times 10^{3}$  \\
    	$\mathcal{F}_6$ & 6th boat & $103.22 \times 10^{3}$  \\
	    \hline
	\end{tabular}
	\label{tab:fishtoat4}
\end{table}

After understanding how the coalition structure and region's fish population change, we come back to the result of total fish caught, either by one individual boat or the entire fleets. Here, we list the results for the structures of grand coalition (Table \ref{tab:fishtoatl}), isolated coalition (Table \ref{tab:fishtoat2}), controlled coalition with fish redistribution (Table \ref{tab:fishtoat3}), and controlled coalition without fish redistribution (Table \ref{tab:fishtoat4}). We find that the results from all these coalition structures are very close, but the grand coalition still has the maximum amount of fish caught. 
\begin{figure}[ht]
\begin{center}
\includegraphics[width=\columnwidth]{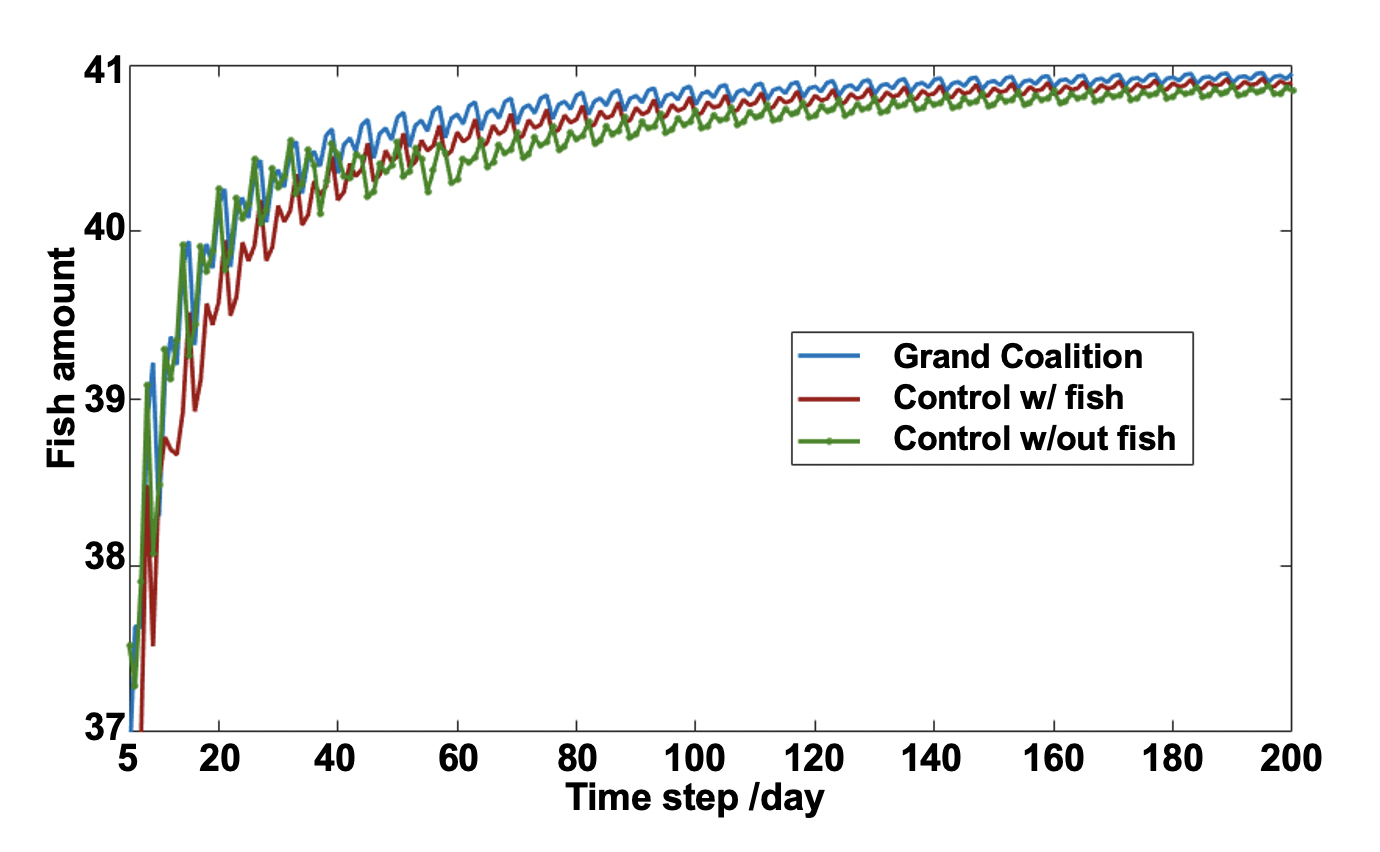}
\caption{Total fish caught of first boat: the comparison between grand coalition, controlled coalition with fish redistribution and controlled coalition without fish redistribution from start to day 200.}
\label{fig:ave}
\end{center}
\end{figure}

Graphically, we show the fish caught of the first boat using different coalition structures in Figure \ref{fig:ave}. Other boats display similar behaviors as to the 1st boat. Still, we observe some differences in terms of fish caught, especially in the first 40 days. When the coalition structure is altered, the amount of fish caught changes accordingly. Also, the periodic solutions of three cases all converge to almost the same point, which means that our controlled methods have good performance compared to social optimum solution. It concludes that even without global information, one agent can still achieve an excellent schedule or plan through communication and cooperation with other agents. 

In total, we can draw the following conclusions from this experiment. First, in our design, the coalition structure becomes stable after some time, and it means that the program can find the steady-state solution in a dynamic problem. Second, the change in coalition structure leads to the change in each region's fish amount and the change of each boat's fish caught. Last, our coalition control method results are close to the result of grand coalition that finds a social optimum solution, whereas our algorithm reaches the equilibrium through the Nash-bargaining process. It suggests that our model is capable of handling more complex problems.

\subsubsection{Heuristic Coalition Control Model}
\begin{figure}[htbp]
\begin{center}
\includegraphics[width=\columnwidth]{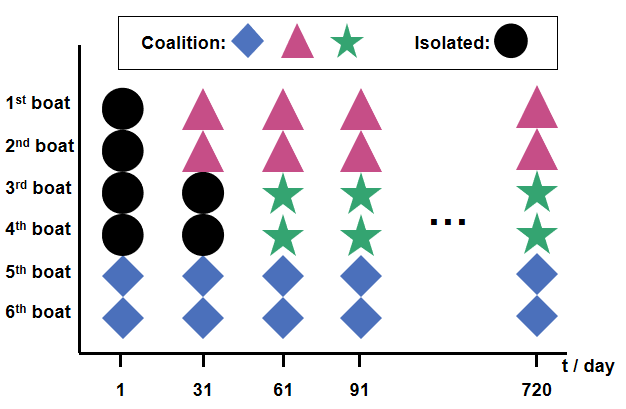}
\caption{How coalition structure changes after each month using heuristic approach. 
Two boats form a coalition each month, and after 3 months, the coalition structure no longer changes.}
\label{fig:hcform}
\end{center}
\end{figure}

We have illustrated our coalition control method, which finds the global optimal solution of the fishing problem. However, due to computing inefficiency, we also develop a method based on hierarchical clustering, which finds the local optimal solution. First, we use the same experimental setup as before, and we set $\mu=0.001$ to keep the real and virtual effort vectors relatively close to each other and $\gamma = 1$ for merging process. The change in coalition structure is shown in Fig.\ref{fig:hcform}. Based on our trade-off parameters and algorithm, boats with similar fishing capability merge to one coalition, and the merging process happens each month. Table \ref{tab:fishhc1} shows the total fish caught of entire fleet and each boat. In Table \ref{tab:fishcom}, we compare the total fish caught of each method used(the controlled method has fish redistribution). It is reasonable that the fish caught from accelerated coalition control method is greater than the result from isolated coalition but less than the results from grand coalition and regular controlled coalition.

 \begin{table}[ht]
	\centering
	\caption{Hierarchical Clustering: The total fish caught of entire fleet and each boat in 2 years/720 time steps.}
	\begin{tabular}{lll}
		Symbol & Description & Fish Value\\
		\hline
		$\mathcal{F}$ & total fish caught & $342.45 \times 10^{3}$  \\
		$\mathcal{F}_1$ & 1st boat & $29.46 \times 10^{3}$  \\	
		$\mathcal{F}_2$ & 2nd boat & $36.82 \times 10^{3}$ \\
    	$\mathcal{F}_3$ & 3rd boat & $44.19 \times 10^{3}$   \\
		$\mathcal{F}_4$ & 4th boat & $55.23 \times 10^{3}$  \\
    	$\mathcal{F}_5$ & 5th boat & $73.65 \times 10^{3}$  \\
    	$\mathcal{F}_6$ & 6th boat & $103.10 \times 10^{3}$  \\
	    \hline
	\end{tabular}
	\label{tab:fishhc1}
\end{table}

 \begin{table}[ht]
	\centering
	\caption{Summary of total fish caught of each coalition control method in 2 years/720 time steps.}
	\begin{tabular}{ll}
		Method & Fish Value\\
		\hline
		Grand & $343.25 \times 10^{3}$  \\
		Isolated & $337.22 \times 10^{3}$  \\	
		Controlled & $343.00 \times 10^{3}$  \\
    	Accelerated & $342.45 \times 10^{3}$  \\
	    \hline
	\end{tabular}
	\label{tab:fishcom}
\end{table}
Also, we have benchmarked the performances of the accelerated method comparing to previous methods. All experiments were performed on PC with Intel i7-7700 CPU having 3.6-GHz clock speed and 56 GB of memory, and all the methods were implemented and tested using MATLAB R2016a. The time recorded is the computing time consumed for one iteration, i.e. one day, and the results are shown in Table \ref{tab:benchmark}. In the table, we keep the same four regions, and then we increase the number of fishing boat from 4 to 24. For our original coalition control method, the time extends dramatically when we increasing the number of fishing boat, and we cannot achieve a solution within reasonable time when the boat number is greater than 18. However, the accelerated method reduces the computing time remarkably, and it suggests that we can apply this method to large scale problems. 

 \begin{table}[ht]
	\centering
	\caption{Running time of different coalitions for one iteration}
	\begin{tabular}{lllll}
		Method & 4$\times$6 & 4$\times$12 & 4$\times$18 & 4$\times$24\\
		\hline
		Grand & 6.1s & 25.8s & 156.1s  & 197.3s\\
		Isolated & 3.0s & 6.6s & 9.6s & 19.1s \\
		Controlled & 382.7s  & 6361.9s  & NA & NA\\	
		Accelerated  & 14.5s & 46.9s & 71.2s & 115.2s \\
	    \hline
	\end{tabular}
	\label{tab:benchmark}
\end{table}

\section{CONCLUSION}

In this paper, we propose coalition control methods based on the information produced by MPC. The methods, including both globally optimal and heuristic approaches, can automatically adjust the coalition structure to optimize a given objective function. Our methods can be applied to the scenarios in which individual and collective interests conflict, and it finds the optimal solution through the Nash-Bargaining process. In the future, we would like to apply our methods to real problems, which may include more complicated mechanisms. As for now, our main concern is to accelerate the coalition process in terms of computing time, even though we know that the future can be hard to predict.

\bibliographystyle{ieeetr}
\bibliography{pasa-sample}

\begin{thebibliography}{10}

\bibitem{optimal_control}
Jasbir~S. Arora.
\newblock {\em Introduction to Optimum Design, 2nd. ed}.
\newblock Academic Press, 5th May 2004.

\bibitem{AFRAM2014343}
Abdul Afram and Farrokh Janabi-Sharifi.
\newblock Theory and applications of hvac control systems -- a review of model
  predictive control (mpc).
\newblock {\em Building and Environment}, 72:343 -- 355, 2014.

\bibitem{8870090}
M.~{Košuda}, J.~{Novotňák}, and M.~{Fiĺko}.
\newblock Energy-oriented trajectory optimization of solar aircraft using
  fmincon function in matlab.
\newblock In {\em Proc. 2019 International Conference on Military Technologies
  (ICMT)}, pages 1--5, 2019.

\bibitem{FELE2014314}
Filiberto Fele, Jos{\'e}~M. Maestre, S.~Mehdy Hashemy, David {Mu{\~n}oz de la
  Pe{\~n}a}, and Eduardo~F. Camacho.
\newblock Coalitional model predictive control of an irrigation canal.
\newblock {\em Journal of Process Control}, 24(4):314 -- 325, 2014.

\bibitem{9143643}
E.~{Masero}, L.~A. {Fletscher}, and J.~M. {Maestre}.
\newblock A coalitional model predictive control approach for heterogeneous
  cellular networks*.
\newblock In {\em Proc. 2020 European Control Conference (ECC)}, pages
  448--453, 2020.

\bibitem{9102355}
P.~{Chanfreut}, J.~M. {Maestre}, and E.~F. {Camacho}.
\newblock Coalitional model predictive control on freeways traffic networks.
\newblock {\em IEEE Transactions on Intelligent Transportation Systems}, pages
  1--12, 2020.

\bibitem{5454336}
B.~{Asadi} and A.~{Vahidi}.
\newblock Predictive cruise control: Utilizing upcoming traffic signal
  information for improving fuel economy and reducing trip time.
\newblock {\em IEEE Transactions on Control Systems Technology},
  19(3):707--714, 2011.

\bibitem{baldiviesomonasterios2020coalitional}
Pablo R~B Monasterios and Paul~A Trodden.
\newblock Coalitional predictive control: consensus-based coalition forming
  with robust regulation.
\newblock 2020.

\bibitem{9029893}
P.~A. {Trodden} and P.~R.~B. {Monasterios}.
\newblock Low-complexity robust decentralized mpc: a foundational algorithm for
  constrained coalitional control.
\newblock In {\em Proc. 2019 IEEE 58th Conference on Decision and Control
  (CDC)}, pages 1089--1095, 2019.

\bibitem{8409399}
F.~J. {Muros}, J.~M. {Maestre}, C.~{Ocampo-Martinez}, E.~{Algaba}, and E.~F.
  {Camacho}.
\newblock A game theoretical randomized method for large-scale systems
  partitioning.
\newblock {\em IEEE Access}, 6:42245--42263, 2018.

\bibitem{soicalop}
T.~{Alpcan} and L.~{Pavel}.
\newblock Nash equilibrium design and optimization.
\newblock In {\em Proc. 2009 International Conference on Game Theory for
  Networks (ICGTN)}, pages 164--170, 2009.

\bibitem{8990420}
M.~{Amirfattahi} and M.~{Haeri}.
\newblock A real-time bargaining-based algorithm for energy trading market in
  smart grid.
\newblock In {\em Proc. 2019 11th International Conference on Electrical and
  Electronics Engineering (ELECO)}, pages 17--21, 2019.

\bibitem{8865184}
D.~{Huang}, Q.~{Yuan}, and X.~{Li}.
\newblock Decentralized flocking of multi-agent system based on mpc with
  obstacle/collision avoidance.
\newblock In {\em Proc. 2019 Chinese Control Conference (CCC)}, pages
  5587--5592, 2019.

\bibitem{economiclmpc}
J.~B. {Rawlings}, D.~{Angeli}, and C.~N. {Bates}.
\newblock Fundamentals of economic model predictive control.
\newblock In {\em Proc. 2012 51st IEEE Conference on Decision and Control
  (CDC)}, pages 3851--3861, 2012.

\bibitem{coalstruct}
H.~{Wu} and S.~{Hu}.
\newblock A coalition structure generation algorithm based on partition
  cardinality structure.
\newblock In {\em Proc. 2010 IEEE International Conference on Intelligent
  Computing and Intelligent Systems (ICICI)}, volume~3, pages 75--78, 2010.

\bibitem{Mayne2014}
D.~Mayne.
\newblock Model predictive control: Recent developments and future promise.
\newblock {\em Autom.}, 50:2967--2986, 2014.

\end{thebibliography}

\end{document}